# Practical and Robust Stenciled Shadow Volumes for Hardware-Accelerated Rendering

Cass Everitt and Mark J. Kilgard
March 12, 2002
NVIDIA Corporation, Copyright 2002
Austin, Texas

## ABSTRACT

Twenty-five years ago, Crow published the shadow volume approach for determining shadowed regions in a scene. A decade ago, Heidmann described a hardware-accelerated stencil buffer-based shadow volume algorithm.

However, hardware-accelerated stenciled shadow volume techniques have not been widely adopted by 3D games and applications due in large part to the lack of robustness of described techniques. This situation persists despite widely available hardware support. Specifically what has been lacking is a technique that robustly handles various "hard" situations created by near or far plane clipping of shadow volumes.

We describe a robust, artifact-free technique for hardware-accelerated rendering of stenciled shadow volumes. Assuming existing hardware, we resolve the issues otherwise caused by shadow volume near and far plane clipping through a combination of (1) placing the conventional far clip plane "at infinity", (2) rasterization with infinite shadow volume polygons via homogeneous coordinates, and (3) adopting a *zfail* stencil-testing scheme. *Depth clamping*, a new rasterization feature provided by NVIDIA's GeForce3 & GeForce4 Ti GPUs, preserves existing depth precision by not requiring the far plane to be placed at infinity. We also propose *two-sided stencil testing* to improve the efficiency of rendering stenciled shadow volumes.

## 1. INTRODUCTION

Crow's shadow volume approach [10] to shadow determination is twenty-five years old. A shadow volume defines a region of space that is in the shadow of a particular occluder given a particular ideal light source. The shadow test determines if a given point being tested is inside the shadow volume of any occluder. Hardware stencil testing provides fast hardware acceleration for shadow determination using shadow volumes. Despite the relative age of the shadow volume approach and the widespread availability of stencil-capable graphics hardware, use of shadow volumes in 3D games and applications is rare.

We believe this situation is due to the lack of a practical and robust algorithm for rendering stenciled shadow volumes. We propose here an algorithm to address this gap. Our algorithm is practical because it requires only features available in OpenGL 1.0. The algorithm is robust because shadow volume scenarios that vexed previous algorithms, such as a light within an open container, are handled automatically and correctly.

We focus on robustly solving the problem of hardware-accelerated stenciled shadow volume rendering for a number of reasons, many noted by other authors [9][10][11][19]:

- Shadow volumes provide omni-directional shadows.

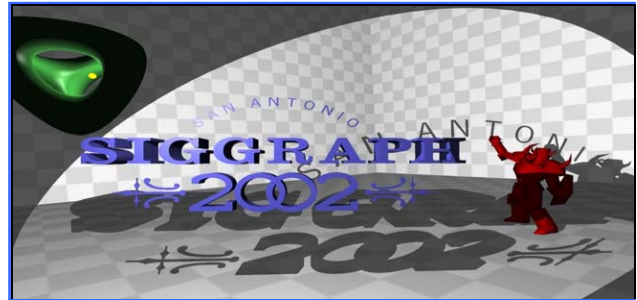

- Shadow volumes automatically handle self-shadowing of objects if implemented correctly.
- Shadow volumes perform shadow determination in window space, resolving shadow boundaries with pixel accuracy (or sub-pixel accuracy when multisampling is available).
- Lastly, the fundamental stencil testing functionality required for hardware-accelerated stenciled shadow volumes is now ubiquitous due to the functionality's standardization by OpenGL 1.0 (1991) and DirectX 6 (1998) respectively. It is near impossible to purchase a new PC in 2002 without stencil testing hardware.

Stenciled shadow volumes have their limitations too. Shadow volumes model ideal light sources so the resulting shadow boundaries lack soft edges. Shadow volume techniques require polygonal models. Unless specially handled, such polygonal models must be closed (2-manifold) and be free of non-planar polygons. Silhouette computations for dynamic scenes can prove expensive. Stenciled shadow volume algorithms are inherently multi-pass. Rendering shadow volumes can consume tremendous amounts of pixel fill rate.

## 2. PREVIOUS WORK

### 2.1 Pre-Stencil Testing Work

Crow [10] first published the shadow volume approach in 1977. Crow recognizes that the front- or back-facing orientations of consistently rendered shadow volume polygons with respect to the viewer indicate enters into and exits out of shadowed regions. Crow also recognizes that some care must be taken to determine if the viewer's eye point is within a shadow volume.

Crow's formulation fundamentally involves walking a pixel's view ray originating at the eye point and counting the number of shadow volume enters and exits encountered prior to the first visible rasterized fragment.

Brotman and Badler [8] in 1984 adapted Crow's shadow volume approach to a software-based, depth-buffered, tiled renderer with deferred shading and support for soft shadows through numerous light sources all casting shadow volumes.

Pixel-Planes [13] in 1985 provides hardware support for shadow volume evaluation. In contrast to Crow's original ray walking approach, the Pixel-Planes algorithm relies on determining if a pixel is within an infinite polyhedron defined by a single occluder triangle plane and its three shadow volume planes. This determination is made for every pixel and for every occluder polygon in the scene. Each "point within a volume" test is computed by evaluating the corresponding set of plane equations.

Pixel-Planes is a unique architecture because the area of a rasterized triangle in pixels does not affect the triangle's rasterization time. Otherwise, the algorithm's evaluation of *every* per-triangle shadow volume plane equation at *every* pixel would be terribly inefficient.

Bergeron [3] in 1986 generalizes Crow's original shadow volume approach. Bergeron explains how to handle open models and models containing non-planar polygons properly. Bergeron explicitly notes the need to close shadow volumes so that a correct initial count of how many shadow volumes the eye is within can be computed.

Fournier and Fussell [12] in 1988 discuss shadow volumes in the context of frame buffer computations. In their computational model, each pixel in a frame buffer maintains a depth value and shadow depth count. Fournier and Fussell's frame buffer computation model lays the theoretical foundations for subsequent hardware stencil buffer-based algorithms.

## 2.2  Stencil Testing-Based Work
### 2.2.1  The Original Approach
Heidmann [14] in 1991 describes an algorithm for using the then-new stencil buffer support of SGI's VGX graphics hardware [1]. Heidmann recognizes the problem of stencil buffer overflows and demonstrates combining contributions from multiple light sources with the accumulation buffer to simulate soft shadows.

Heidmann's approach is a multi-pass rendering algorithm. First, the color, depth, and stencil buffers are cleared. Second, the scene is drawn with only ambient and emissive lighting contributions and using depth testing for visibility determination. Now the color and depth buffers contain the color and depth values for the closest fragment rendered at each pixel. Then shadow volume polygons are rendered into the scene but just updating stencil.

Front-facing polygons update the frame buffer with the following OpenGL per-fragment operations (for brevity, we drop the `gl` and `GL` prefixes for OpenGL commands and tokens):

```
Enable(CULL_FACE);            // Face culling enabled
CullFace(BACK);               // to eliminate back faces
ColorMask(0,0,0,0);           // Disable color buffer writes
DepthMask(0);                 // Disable depth buffer writes
StencilMask(~0);              // Enable stencil writes
Enable(DEPTH_TEST);           // Depth test enabled
DepthFunc(LEQUAL);            // less than or equal
Enable(STENCIL_TEST);         // Stencil test enabled
StencilFunc(ALWAYS,0,~0);     // always pass
StencilOp(KEEP,KEEP,INCR);    // increment on zpass
```

Similarly, back-facing polygons update the frame buffer with the following OpenGL state modifications:

```
CullFace(FRONT);              // Now eliminate front faces
StencilOp(KEEP,KEEP,DECR);    // Now decrement on zpass
```

Heidmann's described algorithm computes the front- or back-facing orientation of shadow volume polygons on the CPU. We note (as have other authors [5][15]) that the shadow volume polygons can be rendered in two passes: first, culling back-facing polygons to increment pixels rasterized by front-facing polygons; second, culling front-facing polygons to decrement pixels rasterized by back-facing polygons. This leverages the graphics hardware's ability to make the face culling determination automatically and minimizes hardware state changes at the cost of rendering the shadow volume polygons twice. Utilizing the hardware's face culling also avoids inconsistencies if the CPU and graphics hardware determine a polygon's orientation differently in razor's edge cases.

After the shadow volume polygons are rendered into the scene, a pixel's stencil value is equal to zero if the light illuminates the pixel and greater than zero if the pixel is shadowed. The scene can then be re-rendered with the appropriate light configured and enabled, with stencil testing enabled to update only pixels with a zero stencil value (meaning the pixel is not shadowed), and "depth equal" depth testing (to update only visible fragments). The light's contribution can be accumulated with either the accumulation buffer or additive blending.

This can be repeated for multiple light sources, clearing the stencil buffer between rendering the shadow volumes and summing the contribution of each light.

### 2.2.2  Near and Far Plane Clipping and Capping
Heidmann fails to mention in his article a problem that seriously undermines the robustness of his approach. With arbitrary scenes, the near and/or far clip planes may (and, in fact, often will) clip the infinite shadow volumes. Each shadow volume is, by construction, a half-space (dividing the entirety of space into the region shadowed by a given occluder and everything else). However, near and far plane clipping can "slice open" an otherwise well-defined half space. Disturbing the shadow volume in this way leads to incorrect shadow depth counting that, in turn, results in glaringly incorrect shadowing.

Diefenbach [11] in 1996 recognized the problem created by near plane clipping for shadow volume rendering. Diefenbach presents a method that he claims works "for any shadow volume geometry from any viewpoint," but the method, in fact, does not work in several cases. Figure 1 illustrates three cases where Diefenbach's method fails.

Another solution to the shadow volume near plane clipping problem mentioned by Diefenbach is capping off the shadow volume's intersection with the near clip plane. Other authors [2][4][9][16][17] have also suggested this approach. The problem with near plane capping of shadow volumes is that it is, as described by Carmack [9], a "fragile" procedure.

Capping involves projecting each occluder's back-facing polygons to the near clip plane. This can be complicated when only one or two of a projected polygon's vertices intersect the near plane and careful plane-plane intersection computations are required in such cases. The capping process is further complicated when a back-facing occluder polygon straddles the near clip plane.

Rendering capping polygons at the near clip plane is difficult because of the razor's edge nature of the near clip plane. If you are not careful, the very near plane you are attempting to cap can clip your capping polygons! Additionally if the capping polygons are not "watertight" (2-manifold) with the shadow volume being capped then rasterization cracks or double hitting of pixels can create shadowing artifacts. These artifacts appear as exceedingly narrow regions of the final scene where areas that clearly should be illuminated are shadowed and vice versa. These artifacts are painfully obvious in animated scenes.

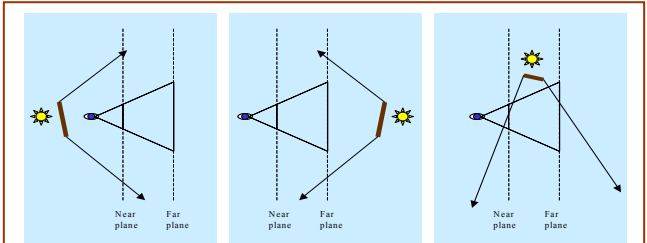

**Figure 1:** Three cases where Deifenbach's capping algorithm fails because some or all pixels requiring capping are covered by neither a front-nor back-facing polygon so Diefenbach's approach cannot correct these pixels.

Watertight capping is non-trivial, particularly if shadow volumes are drawn using object-space geometry so that fast dedicated vertex transformation hardware can be exploited. Kilgard [16] proposes creating a "near plane ledge" whereby closed capping polygons can be rendered in a way that avoids clipping by the near clip plane even when rendering object-space shadow volume geometry. This approach cedes a small amount of depth buffer precision for the ledge. Additionally shadow volume capping polygons must be rendered twice, incrementing front-faces and decrementing back-facing geometry because the orientation (front- or back-facing) of a polygon can occasionally flip when a polygon of nearly zero area in window space is transformed from object space to window space due to floating-point numerics. Otherwise, shadow artifacts result.

Even when done carefully, shadow volume near plane capping is treacherous because of the fragile nature of required ray-plane intersections and the inability to guarantee identical and bit-exact CPU and GPU floating-point computations. In any case, capping computations burden the CPU with an expensive task that our algorithm obviates.

### 2.2.3 Zpass vs. Zfail Stenciled Shadow Volumes

The conventional stenciled shadow volume formulation is to increment and decrement the shadow depth count for front- and back-facing polygons respectively when the depth test passes. Bilodeau [5] in 1999 noted that reversing the depth comparison works too. Another version of this alternative formulation is to decrement and increment the shadow depth count for front- and back-facing polygons respectively depth test fails (without reversing the comparison).

Carmack [9] in 2000 realized the equivalence of the two formulations because they both achieve the *same* result, if in the "depth test fail" formulation, the shadow volume is "closed off" at both ends (rather than being open at the ends). Compare the following OpenGL rendering state modifications with the settings for conventional shadow volume rendering in section 2.2.1.

Front-facing shadow volume rendering configuration:

    `StencilOp(KEEP,DECR,KEEP);` // *decrement* on *zfail*

Back-facing configuration:

    `StencilOp(KEEP,INCR,KEEP);` // *increment* on *zfail*

What Carmack describes is projecting back faces, with respect to the light source, some large but finite distance (importantly, still within the far clip plane) and also treating the front faces of the occluder, again with respect to the light source, as a part of the shadow volume boundary too. This still has a problem because when a light source is arbitrarily close to a single occluder polygon, any finite distance used to project out the back faces of the occluder to close off the shadow volume may not extend far enough to ensure that objects beyond the occluder are properly shadowed.

Still Carmack's insight is fundamental to our new algorithm. We call Bilodeau and Carmack's approach *zfail* stenciled shadow volume rendering because the stencil increment and decrement operations occur when the depth test fails rather than when it passes. We call the conventional approach *zpass* rendering.

One way to think of the *zfail* formulation, in contrast to the *zpass* formulation, is that the *zfail* version counts shadow volume intersections from the opposite direction. The *zpass* formulation counts shadow volume enters and exits along each pixel's view ray between the eye point and the first visible rasterized fragment. Technically due to near plane clipping, the counting occurs only between the ray's intersection point with the near clip plane and the first visible rasterized fragment. The objective of shadow volume capping is to introduce sufficient shadow volume enters so that the eye can always be considered "out of shadow" so the stencil count can reflect the true absolute shadow depth of the first visible rasterized fragment.

The *zfail* formulation instead counts shadow volume enters and exits along each pixel's view ray between *infinity* and the first visible rasterized fragment. Technically due to far plane clipping, the counting occurs only between the ray's intersection with the far plane and the first visible fragment. By capping the open end of the shadow volume at or before the far clip plane, we can force the idea that *infinity* is always outside of the shadow volume.

## 3. OUR ALGORITHM
### 3.1 Requirements
For our algorithm to operate robustly, we require the following:

- Models for occluding objects must be composed of triangles only (avoiding non-planar polygons), be closed (2-manifold), and have a consistent winding order for triangles within the model. Homogeneous object coordinates are permitted, assuming w≥0.
- Light sources must be ideal points. Homogeneous light positions (w≥0) allow both positional and directional lights.
- Connectivity information for occluding models must be available so that silhouette edges with respect to a light position can be determined at shadow volume construction time.
- The projection matrix must be perspective, not orthographic.
- Functionality available in OpenGL 1.0 [18] and DirectX 6: transformation and clipping of homogeneous positions; front and back face culling; masking color and depth buffer writes; depth buffering; and stencil-testing support.
- The renderer must support $N$ bits of stencil buffer precision, where $2^N$ is greater than the maximum shadow depth count ever encountered during the processing of a given scene.

This requirement is scene dependent, but 8 bits of stencil buffer precision (typical for most hardware today) is reasonable for typical scenes.

- The renderer must guarantee "watertight" rasterization (no double hitting of pixels or missed pixels along shared edges of rasterized triangles).

Support for non-planar polygons and open models can be achieved using special case handling along the lines described by Bergeron [3].

## 3.2 Approach

We developed our algorithm by methodically addressing the fundamental limitations of the conventional stenciled shadow volume approach. We combine (1) placing the conventional far clip plane "at infinity"; (2) rasterizing infinite (but fully closed) shadow volume polygons via homogeneous coordinates; and (3) adopting the *zfail* stencil-testing scheme.

This is sufficient to render shadow volumes robustly because it avoids the problems created by the far clip plane "slicing open" the shadow volume. The shadow volumes we construct project "all the way to infinity" through the use of homogeneous coordinates to represent the shadow volume's infinite back projection. Importantly, though our shadow volume geometry is infinite, it is also fully closed. The far clip plane, in eye-space, is infinitely far away so it is impossible for any of the shadow volume geometry to be clipped by it.

By using the *zfail* stencil-testing scheme, we can always assume that infinity is "beyond" all closed shadow volumes if we, in fact, close off our shadow volumes at infinity. This means the shadow depth count can always start from zero for every pixel. We need not worry about the shadow volume being clipped by the near clip plane since we are counting shadow volume enters and exits from infinity, rather than from the eye, due to *zfail* stencil-testing. No fragile capping is required so our algorithm is both robust and automatic.

### 3.2.1 Far Plane at Infinity

The standard perspective formulation of the projection matrix used to transform eye-space coordinates to clip space in OpenGL (see `glFrustum` [18]) is

$$\mathbf{P} = \begin{bmatrix} \frac{2 \times Near}{Right - Left} & 0 & \frac{Right + Left}{Right - Left} & 0 \\ 0 & \frac{2 \times Near}{Top - Bottom} & \frac{Top + Bottom}{Top - Bottom} & 0 \\ 0 & 0 & -\frac{Far + Near}{Far - Near} & -\frac{2 \times Far \times Near}{Far - Near} \\ 0 & 0 & -1 & 0 \end{bmatrix}$$

where *Near* and *Far* are the respective distances from the viewer to the near and far clip planes in eye-space.

**P** is used to transform eye-space positions to clip-space positions:

$$\begin{bmatrix} x_c & y_c & z_c & w_c \end{bmatrix}^T = \mathbf{P} \begin{bmatrix} x_e & y_e & z_e & w_e \end{bmatrix}^T$$

We are interested in avoiding far plane clipping so we only concern ourselves with the third and fourth row of **P** used to compute clip-space $z_c$ and $w_c$. Regions of an assembled polygon with interpolated clip coordinates outside $-w_c \leq z_c \leq w_c$ are clipped by the near and far clip planes.

We consider the limit of **P** as the far clip plane distance is driven to infinity (this is not novel; Blinn [7] mentions the idea):

$$\lim_{Far \to \infty} \mathbf{P} = \mathbf{P_{inf}} = \begin{bmatrix} \frac{2 \times Near}{Right - Left} & 0 & \frac{Right + Left}{Right - Left} & 0 \\ 0 & \frac{2 \times Near}{Top - Bottom} & \frac{Top + Bottom}{Top - Bottom} & 0 \\ 0 & 0 & -1 & -2 \times Near \\ 0 & 0 & -1 & 0 \end{bmatrix}$$

The first, second, and fourth rows of $\mathbf{P_{inf}}$ are the same as **P**; only the third row changes. There is no longer a *Far* distance.

A vertex that is an infinite distance from the viewer is represented in homogeneous coordinates with a zero $w_e$ coordinate. If the vertex is transformed into clip space using $\mathbf{P_{inf}}$, assuming the vertex is in front of the eye, meaning that $z_e$ is negative (the OpenGL convention), then $w_c = z_c$ so this transformed vertex is *not* clipped by the far plane. Moreover, its non-homogeneous depth $z_c/w_c$ must be 1.0, generating the maximum possible depth value.

It may be surprising, but positioning the far clip plane at infinity typically reduces the depth buffer precision only marginally. Consider how much we would need to shrink our window coordinates so we can represent within the depth buffer an infinite eye-space distance in front of the viewer. The projection **P** transforms (0,0,-1,0) in eye-space (effectively, an infinite distance in front of the viewer) to the window-space depth *Far*/(*Far-Near*). The largest window coordinate representable in the depth buffer is 1 so we must scale *Far*/(*Far-Near*) by its reciprocal to "fit" infinity in the depth buffer. This scale factor is (*Far-Near*)/*Far* and is very close to 1 if *Far* is many times larger than *Near* which is typical.

Said another way, using $\mathbf{P_{inf}}$ instead of **P** only compresses the depth buffer precision slightly in typical scenarios. For example, if *Near* and *Far* are 1 and 100, then the depth buffer's precision must be squeezed by just 1% to represent an infinite distance in front of the viewer.

### 3.2.2 Infinite Shadow Volume Polygons

We assume that given a light source position and a closed model with its edge-connectivity, we can determine the subset of *possible silhouette* edges for the model. A possible silhouette edge is an edge shared by two triangles in a model where one of the two triangles faces a given light while the other triangle faces away from the light.

We call these edges "possible silhouette" edges rather than just silhouette edges because these edges are not necessarily boundaries between shadowed and illuminated regions as implied by the conventional meaning of silhouette. It is possible that an edge is an actual silhouette edge, but it is also possible that the edge is itself in shadow.

Assume we have computed the plane equations in the form $Ax+By+Cz+Dw=0$ for every triangle in a given model. The plane equation coefficients must be computed using a vertex ordering consistent with the winding order shared by all the triangles in the model such that $Ax+By+Cz+Dw$ is non-negative when a point $(x,y,z,w)$ is on the front-facing side of the triangle's plane. Assume we also know the light's homogeneous position $L$ in the coordinate space matching the plane equations. For each triangle, evaluate $d=AL_x+BL_y+CL_z+DL_w$ for the triangle's plane equation coefficients and the light's position. If $d$ is negative, then the triangle is back-facing with respect to *L*; otherwise the triangle is front-facing with respect to *L*. Any edge shared by two triangles with one triangle front-facing and the other back-facing is a possible silhouette edge.

To close a shadow volume completely, we must combine three sets of polygons: (1) all of the possible silhouette polygon edges extruded to infinity away from the light; (2) all of the occluder's back-facing triangles, with respect to *L*, projected away from the light to infinity; and (3) all of the occluder's front-facing triangles with respect to *L*.

Each possible silhouette edge has two vertices *A* and *B*, represented as homogeneous coordinates and ordered based on the front-facing triangle's vertex order. The shadow volume extrusion polygon for this possible silhouette is formed by the edge and its projection to infinity away from the light. The resulting quad consists of the following four vertices:

$$\langle B_x, B_y, B_z, B_w \rangle$$
$$\langle A_x, A_y, A_z, A_w \rangle$$
$$\langle A_x L_w - L_x A_w, A_y L_w - L_y A_w, A_z L_w - L_z A_w, 0 \rangle$$
$$\langle B_x L_w - L_x B_w, B_y L_w - L_y B_w, B_z L_w - L_z B_w, 0 \rangle$$

The last two vertices are the homogeneous vector differences of *A-L* and *B-L*. These vertices represent directions heading away from the light, explaining why they have *w* coordinate values of zero. We do assume $A_w \geq 0$, $B_w \geq 0$, $L_w \geq 0$, etc.

When we use a perspective transform of the form $P_{inf}$, we can render shadow volume polygons without the possibility that the far plane will clip these polygons.

For each back-facing occluder triangle, its respective triangle projected to infinity is the triangle formed by the following three vertices:

$$\langle A_x L_w - L_x A_w, A_y L_w - L_y A_w, A_z L_w - L_z A_w, 0 \rangle$$
$$\langle B_x L_w - L_x B_w, B_y L_w - L_y B_w, B_z L_w - L_z B_w, 0 \rangle$$
$$\langle C_x L_w - L_x C_w, C_y L_w - L_y C_w, C_z L_w - L_z C_w, 0 \rangle$$

where *A*, *B*, and *C* are each back-facing occluder triangle's three vertices (in the triangle's vertex order).

The front-facing polygons with respect to *L* are straightforward. Given each triangle's three vertices *A*, *B*, and *C* (in the triangle's vertex order), the triangle is formed by the vertices:

$$\langle A_x, A_y, A_z, A_w \rangle$$
$$\langle B_x, B_y, B_z, B_w \rangle$$
$$\langle C_x, C_y, C_z, C_w \rangle$$

Together, these three sets of triangles form the closed geometry of an occluder's shadow volume with respect to the given light.

## 3.3 Rendering Procedure

Now we sketch the complete rendering procedure to render shadows with our technique. Pseudo-code with OpenGL commands is provided to make the procedure more concrete.

1. Clear the depth buffer to 1.0; clear the color buffer.
   `Clear(DEPTH_BUFFER_BIT | COLOR_BUFFER_BIT);`

2. Load the projection with $P_{inf}$ given the aspect ratio, field of view, and near clip plane distance in eye-space.
   ```
   float Pinf[4][4];
   Pinf[1][0] = Pinf[2][0] = Pinf[3][0] = Pinf[0][1] =
   Pinf[2][1] = Pinf[3][1] = Pinf[0][2] = Pinf[1][2] =
   Pinf[0][3] = Pinf[1][3] = Pinf[3][3] = 0;
   Pinf[0][0] = cotangent(fieldOfView)/aspectRatio;
   Pinf[1][1] = cotangent(fieldOfView);
   Pinf[3][2] = -2*near; Pinf[2][2] = Pinf[2][3] = -1;
   MatrixMode(PROJECTION); LoadMatrixf(&Pinf[0][0]);
   ```

3. Load the modelview matrix with the scene's viewing transform.
   `MatrixMode(MODELVIEW); loadCurrentViewTransform();`

4. Render the scene with depth testing, back-face culling, and all light sources disabled (ambient & emissive illumination only).
   ```
   Enable(DEPTH_TEST); DepthFunc(LESS);
   Enable(CULL_FACE); CullFace(BACK);
   Enable(LIGHTING); Disable(LIGHT0);
   LightModelfv(LIGHT_MODEL_AMBIENT, &globalAmbient);
   drawScene();
   ```

5. Disable depth writes, enable additive blending, and set the global ambient light contribution to zero (and zero any emissive contribution if present).
   ```
   DepthMask(0);
   Enable(BLEND); BlendFunc(ONE,ONE);
   LightModelfv(LIGHT_MODEL_AMBIENT, &zero);
   ```

6. For each light source:
   A. Clear the stencil buffer to zero.
      `Clear(STENCIL_BUFFER_BIT);`
   B. Disable color buffer writes and enable stencil testing with the *always* stencil function and writing stencil..
      ```
      ColorMask(0,0,0,0);
      Enable(STENCIL_TEST);
      StencilFunc(ALWAYS,0,~0); StencilMask(~0);
      ```
   C. For each occluder:
      a. Determine whether each triangle in the occluder's model is front- or back-facing with respect to the light's position. Update `triList[].backfacing`.
      b. Configure *zfail* stencil testing to increment stencil for back-facing polygons that fail the depth test.
         `CullFace(FRONT); StencilOp(KEEP,INCR,KEEP);`
      c. Render all possible silhouette edges as quads that project from the edge away from the light to infinity.

```
Vert L = currentLightPosition;
Begin(QUADS);
  for (int i=0; i<numTris; i++) // for each triangle
    // if triangle is front-facing with respect to the light
    if (triList[i].backFacing==0)
      for (int j=0; j<3; j++) // for each triangle edge
        // if adjacent triangle is back-facing
        //   with respect to the light
        if (triList[triList[i].adjacent[j]].backFacing) {
          // found possible silhouette edge
          Vert A = triList[i].v[j];
          Vert B = triList[i].v[(j+1) % 3]; // next vertex
```

```
      Vertex4f(B.x,B.y,B.z,B.w);
      Vertex4f(A.x,A.y,A.z,A.w);
      Vertex4f(A.x*L.w-L.x*A.w,
               A.y*L.w-L.y*A.w,
               A.z*L.w-L.z*A.w, 0); // infinite
      Vertex4f(B.x*L.w-L.x*B.w,
               B.y*L.w-L.y*B.w,
               B.z*L.w-L.z*B.w, 0); // infinite
    }
End(); // quads
```

d. Specially render all occluder triangles. Project and render back facing triangles away from the light to infinity. Render front-facing triangles directly.

```
#define V triList[i].v[j] // macro used in Vertex4f calls
Begin(TRIANGLES);
  for (int i=0; i<numTris; i++) // for each triangle
    // if triangle is back-facing with respect to the light
    if (triList[i].backFacing)
      for (int j=0; j<3; j++) // for each triangle vertex
        Vertex4f(V.x*L.w-L.x*V.w, V.y*L.w-L.y*V.w,
                 V.z*L.w-L.z*V.w, 0); // infinite
    else
      for (int j=0; j<3; j++) // for each triangle vertex
        Vertex4f(V.x,V.y,V.z,V.w);
End(); // triangles
```

e. Configure *zfail* stencil testing to decrement stencil for front-facing polygons that fail the depth test.
```
CullFace(BACK);  StencilOp(KEEP,DECR,KEEP);
```

f. Repeat steps (c) and (d) above, this time rendering front facing polygons rather than back facing ones.

D. Position and enable the current light (and otherwise configure the light's attenuation, color, etc.).
```
Enable(LIGHT0);
Lightfv(LIGHT0, POSITION, ¤tLightPosition.x);
```

E. Set stencil testing to render only pixels with a zero stencil value, i.e., visible fragments illuminated by the current light. Use *equal* depth testing to update only the visible fragments, and then, increment stencil to avoid double blending. Re-enable color buffer writes again.
```
StencilFunc(EQUAL, 0, ~0); StencilOp(KEEP,KEEP,INCR);
DepthFunc(EQUAL);  ColorMask(1,1,1,1);
```

F. Re-draw the scene to add the contribution of the current light to illuminated (non-shadowed) regions of the scene.
```
drawScene();
```

G. Restore the depth test to *less*.
```
DepthFunc(LESS);
```

7. Disable blending and stencil testing; re-enable depth writes.
```
Disable(BLEND);  Disable(STENCIL_TEST);  DepthMask(1);
```

## 3.4 Optimizations
Possible silhouette edges form closed loops. If a loop of possible silhouette edges is identified, then sending QUAD_STRIP primitives (2 vertices/projected quad), rather than independent quads (4 vertices/projected quad) will reduce the per-vertex transformation overhead per shadow volume quad. Similarly, the independent triangle rendering used for capping the shadow volumes can be optimized for rendering as triangle strips or indexed triangles.

The INCR *zpass* stencil operation in step 6.E avoids the double blending of lighting contributions in the usually quite rare circumstance when two fragments alias to the exact same pixel location and depth value. Using the KEEP *zpass* stencil operation instead can avoid usually unnecessary stencil buffer writes, improving rendering performance in situations where double blending is deemed unlikely.

In the case of a directional light, all the vertices of a possible silhouette edge loop project to the same point at infinity. In this case, a TRIANGLE_FAN primitive can render these polygons extremely efficiently (1 vertex/projected triangle).

If the application determines that the shadow volume geometry for a silhouette edge loop will never pierce or otherwise require capping of the near clip plane's visible region, *zpass* shadow volume rendering can be used instead of *zfail* rendering. The *zpass* formulation is advantageous in this context because it does not require the rendering of any capping triangles. Mixing the *zpass* and *zfail* shadow volume stencil testing formulations for different silhouette edge loops does not affect the net shadow depth count as long as each particular loop uses a single formulation.

Shadow volume geometry can be re-used from frame to frame for any light and occluder that have not changed their geometric relationship to each other.

## 4. IMPROVED HARDWARE SUPPORT
### 4.1 Wrapping Stencil Arithmetic
DirectX 6 and the OpenGL *EXT_stencil_wrap* extension provide two additional *increment wrap* and *decrement wrap* stencil operations that use modulo, rather than saturation, arithmetic. These operations reduce the likelihood of incorrect shadow results due to an increment operation saturating a stencil value's shadow depth count. Using the wrapping operations with an *N*-bit stencil buffer, there remains a remote possibility that a net $2^N$ increments (or a multiple of) may alias with the unshadowed zero stencil value and lead to incorrect shadows, but in practice, particularly with an 8-bit stencil buffer, this is quite unlikely.

### 4.2 Depth Clamping
NVIDIA's GeForce3 and GeForce4 Ti GPUs support *depth clamping* via the *NV_depth_clamp* OpenGL extension. When enabled, depth clamping disables the near and far clip planes for rasterizing geometric primitives. Instead, a fragment's window-space depth value is clamped to the range [min(*zn*,*zf*),max(*zn*,*zf*)] where *zn* and *zf* are the near and far depth range values. Additionally when depth clamping is enabled, no fragments with non-positive $w_c$ are generated.

With depth clamping support, a conventional projection matrix with a finite far clip plane distance can be used rather than the $P_{inf}$ form. The only required modification to our algorithm is enabling DEPTH_CLAMP_NV during the rendering of the shadow volume geometry.

Depth clamping recovers the depth precision (admittedly quite marginal) lost due to the use of a $P_{inf}$ projection matrix. More significantly, depth clamping generalizes our algorithm so it works with orthographic, not just perspective, projections.

### 4.3 Two-Sided Stencil Testing

We propose *two-sided stencil testing*, a new stencil functionality that uses distinct front- and back-facing stencil state when enabled. Front-facing primitives use the front-facing stencil state for their stencil operation while back-facing primitives use the back-facing state. With two-sided stencil testing, shadow volume geometry need only be rendered once, rather than twice.

Two-sided stencil testing generates the same number of stencil buffer updates as the two-pass approach so in fill-limited shadow volume rendering situations, the advantage of a single pass is marginal. However, pipeline bubbles due to repeated all front-facing or all back-facing shadow volumes lead to inefficiencies using two passes. Perhaps more importantly, two-sided stencil testing reduces the CPU overhead in the driver by sending shadow volume polygon geometry only once.

Because stencil increments and decrements are intermixed with two-sided stencil testing, the wrapping versions of these operations are mandatory.

### 5. EXAMPLES

Figures 2 through 5 show several examples of our algorithm.

### 6. FUTURE WORK

Because of the extremely scene-dependent nature of shadow volume rendering performance and space constraints here, we defer thorough performance evaluation of our technique. Still we are happy to report that our rendering examples, including examples that seek to mimic the animated behavior of a sophisticated 3D game (see Figure 4), achieve real-time rates on current PC graphics hardware.

Yet naïve rendering with stenciled shadow volumes consumes tremendous amounts of stencil fill rate. We expect effective shadow volume culling schemes will be required to achieve consistent interactive rendering rates for complex shadowed scenes. Portal, BSP, occlusion, and view frustum culling techniques can all improve performance by avoiding the rendering of unnecessary shadow volumes. Additional performance scaling will be through faster and cleverer hardware designs that are better tuned for rendering workloads including stenciled shadow volumes.

Future graphics hardware will support more higher-order graphics primitives beyond triangles. Combining higher-order hardware primitives with shadow volumes requires automatic generation of shadow volumes in hardware. Two-sided stencil testing will be vital since it only requires one rendering of automatically generated shadow volume geometry. Automatic generation of shadow volumes will also relieve the CPU of this chore.

### 7. CONCLUSIONS

Our stenciled shadow volume algorithm is robust, straightforward, and requires hardware functionality that is ubiquitous today. We believe this will provide the opportunity for 3D games and applications to integrate shadow volumes into their basic rendering repertoire. The algorithm we developed is the result of careful integration of known, but not previously integrated, techniques to address methodically the shortcomings of existing shadow volume techniques caused by near and/or far plane clipping.

### 8. ACKNOWLEDGEMENTS

We are grateful to John Carmack, Matt Craighead, Eric Haines, and Matt Papakipos for fruitful discussions, Steve Burke for modeling the SIGGRAPH logo and Loop subdivision surface for us; and James Green, Brian Collins, Rich B, and Stecki for designing and animating the Quake2 models that we picture.

### REFERENCES

[1] Kurt Akeley and James Foran, "Apparatus and method for controlling storage of display information in a computer system," *US Patent 5,394,170*, filed Dec. 15, 1992, assigned Feb. 28, 1995.

[2] Harlen Costa Batagelo and Ilaim Costa Junior, "Real-Time Shadow Generation Using BSP Trees and Stencil Buffers," *XII Brazilian Symposium on Computer Graphics and Image Processing*, Campinas, Brazil, Oct. 1999, pp. 93-102.

[3] Philippe Bergeron, "A General Version of Crow's Shadow Volumes," *IEEE Computer Graphics and Applications*, Sept. 1986, pp. 17-28.

[4] Jason Bestimt and Bryant Freitag, "Real-Time Shadow Casting Using Shadow Volumes," Gamasutra.com web site, Nov. 15, 1999.

[5] Bill Bilodeau and Mike Songy, Creative Labs sponsored game developer conference, unpublished slides, Los Angeles, May 1999.

[6] David Blythe, Tom McReynolds, et.al., "Shadow Volumes," *Program with OpenGL: Advanced Rendering*, SIGGRAPH course notes, 1996.

[7] Jim Blinn, "A Trip Down the Graphics Pipeline: The Homogeneous Perspective Transform," *IEEE Computer Graphics and Applications*," May 1993, pp. 75-88.

[8] Lynne Brotman and Norman Badler, "Generating Soft Shadows with a Depth Buffer Algorithm," *IEEE Computer Graphics and Applications*, Oct. 1984, pp. 5-12.

[9] John Carmack, unpublished correspondence, early 2000.

[10] Frank Crow, "Shadow Algorithms for Computer Graphics," *Proceedings of SIGGRAPH*, 1977, pp. 242-248.

[11] Paul Diefenbach, *Multi-pass Pipeline Rendering: Interaction and Realism through Hardware Provisions,* Ph.D. thesis, University of Pennsylvania, tech report MS-CIS-96-26, 1996.

[12] Alain Fournier and Donald Fussell, "On the Power of the Frame Buffer," *ACM Transactions on Graphics*, April 1988, 103-128.

[13] Henry Fuchs, Jack Goldfeather, Jeff Hultquist, Susan, Spach, John Austin, Frederick Brooks, John Eyles, and John Poulton, "Fast Spheres, Shadows, Textures, Transparencies, and Image Enhancements in Pixel-Planes," *Proceedings of SIGGRAPH*, 1985, pp. 111-120.

[14] Tim Heidmann, "Real Shadows Real Time", *IRIS Universe,* Number 18, 1991, pp. 28-31.

[15] Mark Kilgard, "Improving Shadows and Reflections via the Stencil Buffer," *Advanced OpenGL Game Development* course notes, Game Developer Conference, March 16, 1999, pp. 204-253.

[16] Mark Kilgard, "Robust Stencil Volumes," CEDEC 2001 presentation, Tokyo, Sept. 4, 2001.

[17] Michael McCool, "Shadow Volume Reconstruction from Depth Maps," *ACM Transactions on Graphics*, Jan. 2001, pp. 1-25.

[18] Mark Segal and Kurt Akeley, *The OpenGL Graphics System: A Specification*, version 1.3, 2001.

[19] Andrew Woo, Pierre Poulin, and Alain Fournier, "A Survey of Shadow Algorithms," *IEEE Computer Graphics and Applications*, Nov. 1990, pp. 13-32.

# Practical and Robust Stenciled Shadow Volumes for Hardware-Accelerated Rendering

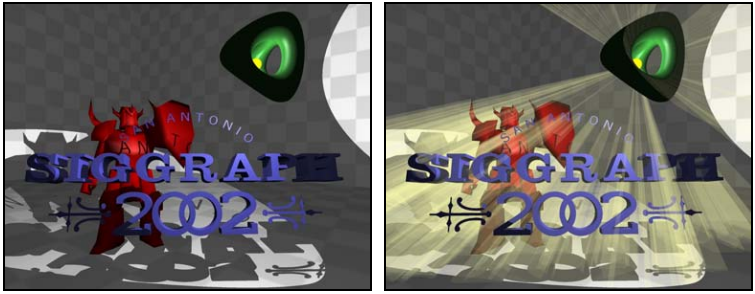

Scene from eye's point of view (**left**) and visualizing shadow volumes (**right**).

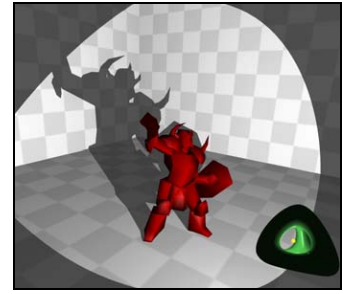

Shadowed scene with the light near the eye and surrounded by a complex surface.

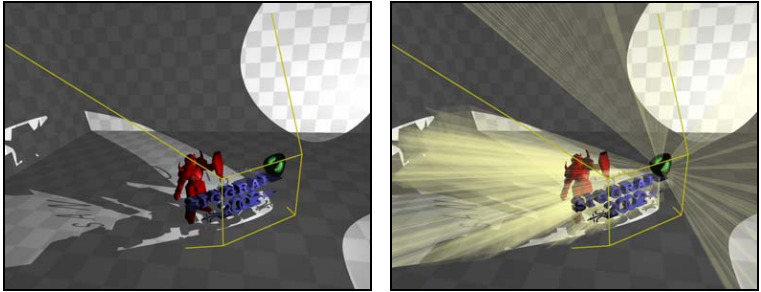

Alternate view of scene showing eye frustum (**left**) and visualizing shadow volumes (**right**).

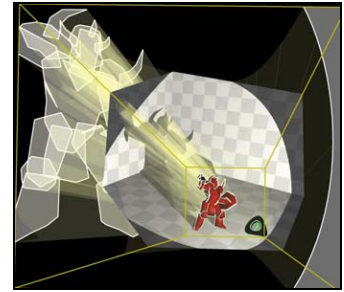

An alternate view of the scene including shadow volumes and silhouette edges with everything outside the eye's infinite frustum clipped away.

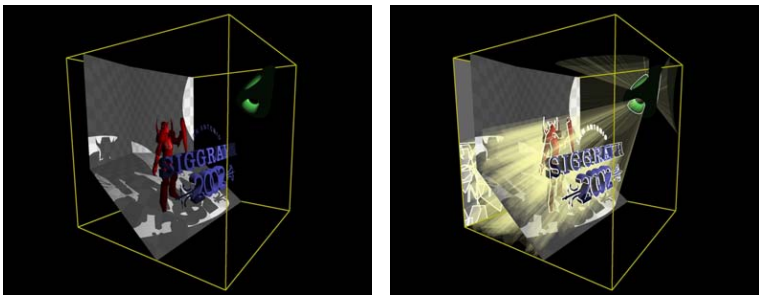

View of scene in eye's clip space (**left**) and visualizing shadow volumes (**right**).

**Figure 2:** These images show a scene with a yellow light source surrounded by a green complex object. This arrangement is a "hard" case for shadow volume rendering. The infinite capping polygons can be seen behind the wall and floor in the bottom right image. All the scenes use a $P_{inf}$ projection matrix.

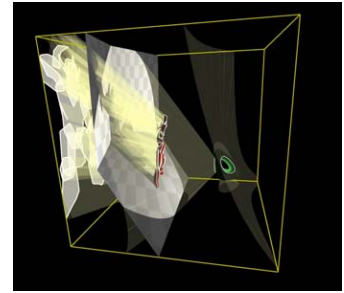

Same as above, except the scene is shown in eye's clip space.

**Figure 3:** These images illustrate the capping at infinity that is required for correct closed shadow computation.

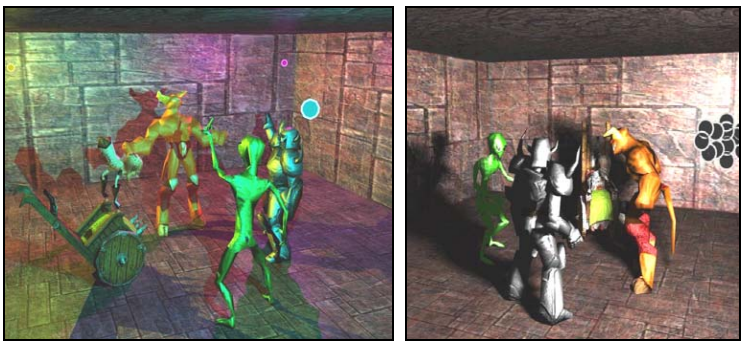

**Figure 4:** Game-like scenes with 3 independent colored light sources (**left**, 34 frames/second on a GeForce4 Ti 4600 at 640x480, 80+ fps for 1 light) and 12 clustered lights to simulate soft shadows (**right**, 8 fps). Characters have diffuse/specular per-pixel bump map shading, correctly shadowed.

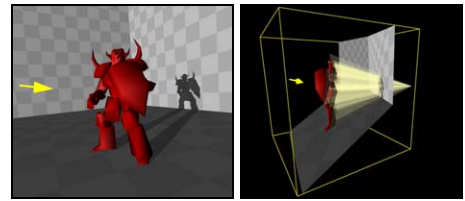

**Figure 5:** Shadowed scene lit by a directional light (**left**) and the corresponding clip-space view with the shadow volume's back projection meeting at infinity on the far clip plane (**right**).